\documentclass[apj]{emulateapj}
\slugcomment{{\sc Accepted to ApJL:} March 17, 2010}

\usepackage{graphicx}
\usepackage{epstopdf}
\usepackage{stmaryrd}
\usepackage{nicefrac}
\usepackage{color}

\newcommand{\etal}{\mbox{et~al.}}

\def\deg      {{\ifmmode^\circ\else$^\circ$\fi}} 

\shorttitle{The Opacity of Galactic Disks at z\,$\sim$\,0.7}
\shortauthors{Sargent et al.}

\begin{document}

\title{The Opacity of Galactic Disks at z\,$\sim$\,0.7}

\author{M.~T. Sargent\altaffilmark{1, 2, $\star$},
C.~M. Carollo\altaffilmark{2},
P. Kampczyk\altaffilmark{2},
S.~J. Lilly\altaffilmark{2},
C. Scarlata\altaffilmark{3},
P. Capak\altaffilmark{4},
O. Ilbert\altaffilmark{5},
A.~M. Koekemoer\altaffilmark{6},
J.-P. Kneib\altaffilmark{5},
A. Leauthaud\altaffilmark{7},
R. Massey\altaffilmark{8},
P.~A. Oesch\altaffilmark{2},
J. Rhodes\altaffilmark{9},
E. Schinnerer\altaffilmark{1},
N. Scoville\altaffilmark{4},
Y. Taniguchi\altaffilmark{10}}

\altaffiltext{1}{Max-Planck-Institut f\"ur Astronomie, K\"onigstuhl 17, D-69117 Heidelberg, Germany}
\altaffiltext{2}{Department of Physics, ETH Zurich, CH-8093 Zurich, Switzerland}
\altaffiltext{3}{{\it Spitzer} Science Center, MC 314-6, California Institute of Technology, Pasadena, CA 91125}
\altaffiltext{4}{California Institute of Technology, MC 105-24, 1200 East California Boulevard, Pasadena, CA 91125, USA}
\altaffiltext{5}{Laboratoire dÕAstrophysique de Marseille, CNRS-Universit\'e dÕAix-Marseille, 38 rue Fr\'ed\'eric Joliot-Curie, 13388 Marseille Cedex 13, France}
\altaffiltext{6}{Space Telescope Science Institute, 3700 San Martin Drive, Baltimore, MD 21218, USA}
\altaffiltext{7}{Physics Division, Lawrence Berkeley National Laboratory, Berkeley, CA 94720, USA}
\altaffiltext{8}{Royal Observatory, Blackford Hill, Edinburgh EH9 3HJ, U.K.}
\altaffiltext{9}{Jet Propulsion Laboratory, California Institute of Technology, 4800 Oak Grove Drive, Pasadena, CA 91109, USA}
\altaffiltext{10}{Research Center for Space and Cosmic Evolution, Ehime University, 2-5 Bunkyo-cho, Matsuyama 790-8577, Japan}

\altaffiltext{$\star$}{E-mail: \texttt{markmr@mpia.de}}

\begin{abstract}
We compare the surface brightness-inclination relation for a sample of COSMOS pure disk galaxies at $z$\,$\sim$\,0.7 with an artificially redshifted sample of SDSS disks well matched to the COSMOS sample in terms of rest-frame photometry and morphology, as well as their selection and analysis.\\
The offset between the average surface brightness of face-on and edge-on disks in the redshifted SDSS sample matches that predicted by measurements of the optical depth of galactic disks in the nearby universe. In contrast, large disks at $z$\,$\sim$\,0.7 have a virtually flat surface brightness-inclination relation, suggesting that they are more opaque than their local counterparts. This could be explained by either an increased amount of optically thick material in disks at higher redshift, or a different spatial distribution of the dust.
\end{abstract}


\section{Introduction}
After a long-lasting controversy on whether galactic disks are optically thick or thin \citep[e.g.][]{disney89, burstein91}, a consensus emerged during the last decade that disks in the local universe ({\it i}\,) behave like optically thick systems as far as integrated photometric properties in the ultraviolet (UV) and visual are concerned \citep[e.g.][]{shao07, driver07, maller09}, while at the same time ({\it ii}\,) peripheral and inter-arm regions in spiral galaxies are more transparent than spiral arms or the core of the disk \citep[e.g.][and references therein]{white00, holwerda05}.\\
Beyond the nearby universe, observational constraints on the extinction in disk galaxies have been hard to obtain \citep[see, e.g., the review by][]{calzetti01} as ({\it i}\,) some of the locally employed measurement techniques can only be gainfully applied over a limited redshift range \citep[e.g.][]{holwerda07}; ({\it ii}\,) the spectral energy distribution (SED) of distant galaxies between the UV and infrared (IR) is often only partially sampled; and ({\it iii}\,) statistical tests -- such as attenuation-inclination relations -- could only be applied to fairly small samples \citep[e.g.][]{lilly98}.\\
Pan-chromatic wide field surveys like COSMOS in principle provide both the necessary multi-wavelength information, as well as the means to select a statistically significant and (morphologically) well-defined sample of disk galaxies thanks to high-resolution HST ({\it Hubble Space Telescope}) imaging. Here we perform a comparative study of the surface brightness-inclination relation in a sample of local and distant ($z$\,$\sim$\,0.7) pure disk galaxies with the aim of testing if their average opacity has evolved in the past 6.5\,Gyr. Our galaxy samples are described in \S\,\ref{sec:samples}, followed by results (\S\,\ref{sec:results}) and conclusions (\S\,\ref{sec:conclu}).

\noindent We adopt a concordance cosmology with $\Omega_m$\,=\,0.25, $\Omega_{\Lambda}+\Omega_m$\,=\,1 and $H_0$\,=\,70\,km\,s$^{-1}$\,Mpc$^{-1}$. Magnitudes are given in the AB-system of \cite{oke74}.

\section{The Disk Samples}
\label{sec:samples}

\subsection{Strategy}
\label{sec:setup}
A study of the opacity of distant disk galaxies that relies on structural information (e.g., surface brightness) requires both high-resolution imaging and a consistent sampling of rest-frame wavelengths. Avoiding so-called `morphological $K$-corrections' is especially important in view of, e.g., size gradients with color \citep[e.g., ][]{dejong96}, and also because unrecognized bulge components can induce spurious signals in attenuation-inclination relations \citep{tuffs04}.\\
The limitation of COSMOS {\it HST} imaging to a single band \citep[the F814W filter; see][]{scoville07} prevents the construction of disk galaxy samples over a continuous redshift range. We hence resort to a comparative study, taking advantage of the fact that the central wavelength of the SDSS $g$-band matches the rest-frame wavelength of objects observed in the F814W filter at redshift z\,$\sim$\,0.7. By artificially redshifting local galaxies to z\,$\sim$\,0.7 \citep[cf.][]{kampczyk07} it becomes possible to directly compare the attenuation properties of distant disks to those in the nearby universe. Moreover, since our COSMOS disks and the redshifted local sample are well matched in terms of both photometric and morphological $K$-corrections, comparisons between the two data sets should eliminate systematic biases.

\subsection{COSMOS Disks at 0.6\,$<$\,{\it z}\,$<$\,0.8}
\label{sec:cosmos}

\subsubsection{Morphological Measurements}
\label{sec:acs}
The {\it z}\,$\sim$\,0.7 COSMOS disks used for the present analysis are selected from a complete sample of 55,324 galaxies with $I$\,$\leq$\,22.5 listed in the Zurich Structure and Morphology Catalog v1.0 (`ZHSM catalog'; Carollo et al., in prep.). Morphological measurements are carried out on {\it HST}/ACS F814W ($I$-band) images with a pixel scale of 0.05$''$/px and resolution of $\sim$0.1$''$ \citep{koekemoer07}. 99.5\% of all catalog entries have a classification as either `early-type', `late-type' or `irregular/peculiar' according to the ZEST algorithm \citep[Zurich Estimator of Structural Types (v1);][]{scarlata07}. Briefly, ZEST performs a principle component (PC) analysis on five non-parametric structural estimators: asymmetry $A$, concentration $C$, Gini coefficient $G$, second-order moment of the brightest 20\% of the pixels associated with an individual galaxy, $M_{20}$, and ellipticity $\epsilon$. Late-type galaxies (ZEST \texttt{CLASS}\,=\,2) are further divided into four bulgeness categories based on the distribution of S\'ersic indices in the different regions of a three-dimensional PC space\footnote{~The ZEST classification scheme reduces the initially five-dimensional space of structural parameters to three dimensions. The axes defining the new metric are linear combinations of $A$, $C$, $G$, $M_{20}$ and $\epsilon$, and correspond to the three dominant principle components. They represent 92\% of the variance in the data set.} occupied by these galaxies.\\
The complete sample has also been modeled with single-component \citet{sersic68} profiles using the \textsc{Gim2d} package \citep{marlsimard98, simard02}. We thus obtain total flux $F_{\rm tot}$; semi-major axis half-light radius $R_{1/2}$; ellipticity $e=1-b/a$ (where $a$ and $b$ are the semi-major and semi-minor axes of the fitted S\'ersic profile); and the value of the S\'ersic index $n$ \citep[for details see][]{sargent07}. For the current analysis we will only use the recovered size and axis ratio, which are essential to estimating the surface brightness and orientation of our disks.\\
We select all COSMOS disk galaxies with a negligibly small bulge component (ZEST \texttt{CLASS}\,=\,2.3; median S\'ersic index $\overline{n}$\,$<$\,1) that are not candidates for being either spurious sources (flags \texttt{JUNK}\,$>$\,0 and \texttt{ACS\_CLEAN}\,=\,0) or stars (flag \texttt{STELLARITY}\,$>$\,0). This results in a sample of 6015 pure disks with $I$\,$\leq$\,22.5 that are not located in an area where the COSMOS multi-wavelength photometry (cf. \S\,\ref{sec:photomcat}) is affected by charge bleeds from saturated stars.

\subsubsection{Multi-Wavelength Photometry \& Redshift Estimates}
\label{sec:photomcat}

The $i^+$-band-selected COSMOS photometry catalog (\citealp{ilbert09}; P. Capak et al. 2010, in prep.) contains $>$500,000 sources with $i$\,$<$\,25, is complete down to this magnitude and provides PSF-matched (FWHM\,=\,1.5$''$) photometry in 30 broad, medium and narrow band filters. The wavelength range covered by these observations extends from 1550\,\AA\ to 8\,$\mu$m \citep{taniguchi07, capak07, sanders07}.\\
We cross-correlated the ZHSM catalog with the photometry catalog in order to obtain multi-wavelength and distance information for our disk galaxy sample. Straightforward positional matching using a search radius of 0.6$''$ resulted in ground-based optical identifications for 1,598 pure disk galaxies in the redshift range of interest, 0.6\,$<$\,{\it z}\,$<$ 0.8. Twenty-three of these were excluded because the association between ACS and ground-based candidates was not unique. Most of these ambiguous matches consist of two ACS objects being blended into a single source in the ground-based imaging.

\noindent Successfully matched disks galaxies generally have a photometric redshift estimated with the code {\it Le Phare} \citep{ilbert09}. The accuracy of the photometric redshifts was calibrated with more than 4000 high-confidence spectroscopic redshifts from zCOSMOS sources \citep{lilly09} with $i^+$\,$\leq$\,22.5. In this magnitude range and at $z$\,$<$\,1.25, \citeauthor{ilbert09} found that photo-{\it z} measurements have a dispersion of $\sigma(\nicefrac{\Delta z}{(1+z)})$\,$\simeq$\,0.007. Based on this statistically expected accuracy of the photometric redshifts, we exclude objects with uncertain photo-{\it z}s. To remain in the sample, the width of the photo-{\it z} probability distribution for a specific source must be $<$2\,$\sigma$ of the dispersion. 30 ($\sim$2\%) of the 1575 disks were rejected due to this criterion, leaving 1545 pure disk galaxies with good quality photometric redshifts 0.6\,$<$\,{\it z}\,$<$\,0.8.

\subsection{SDSS Reference Sample}
\label{sec:SDSS}
\citet{kampczyk07} use SDSS $g$-band images of a volume-limited sample of local disk galaxies (median redshift $z$\,$\sim$\,0.02) to simulate the appearance of these galaxies in COSMOS ACS $I$-band images if redshifted to $z$\,$\sim$\,0.7. As mentioned in \S\,\ref{sec:setup} the ACS F814W filter almost perfectly samples the rest-frame SDSS $g$-band at this redshift. Furthermore, \citet{kampczyk07} account for effects of band-pass shifting, cosmological surface brightness dimming, changing spatial resolution and increased noise. The simulations used here do not include luminosity evolution as its implementation is not straightforward (e.g. non-uniform brightening of galaxy pixels as conceivably caused by a dust component of unknown geometry and composition).\\
The SDSS$_{z=0.7}$ galaxies were processed in exactly the same way as the real COSMOS data. This not only applies to structural measurements and fits to the SED, but also to source detection with SExtractor \citep{bert96} following the insertion of the simulated galaxies into COSMOS ACS pointings. In particular, the same SExtractor configuration parameters used to generate the original list of COSMOS ACS detections in \citet{leauthaud07} were adopted for the extraction of the SDSS$_{z=0.7}$ objects. About 6\% of the SDSS$_{z=0.7}$ galaxies were not recovered by SExtractor. The success rate of the morphological measurements was 100\%. We thus obtain a comparison sample of 75 pure disk galaxies that, after redshifting, are brighter than $I$\,=\,22.5.

\section{Results}
\label{sec:results}
In the following we consider the surface brightness-inclination relation in the rest-frame $B$-band. This has the twofold advantage of bordering on the SED region ($g$-band) sampled by our SDSS and COSMOS imaging, and of providing results that are readily comparable with predictions and previous findings in the literature.

\noindent The average surface brightness is obtained by distributing half (hence the second term in eq. \ref{eq:mufctofflux}) of the total $I$-band flux over an ellipse with a semi-major axis $R_{1/2}$ (in arcseconds). The semi-minor axis follows from the \textsc{Gim2d} value of the axis ratio:
\begin{eqnarray}
\mu(B) = I&+&2.5\,\textrm{log}\left(2\right)+2.5\,\textrm{log}\left(\pi R_{1/2}^2\right)\\ \nonumber
&+&2.5\,\textrm{log}\left(\nicefrac{b}{a}\right)-10\,\textrm{log}\left(1+z\right)-K_{F814W,B}(z)~.
\label{eq:mufctofflux}
\end{eqnarray}
The $K$-correction $K_{F814W,B}(z)$ is determined using the best-fit SED in the template library of the ZEBRA package \citep{feldmann06}. The relation between $\mu(B)$ and that of the system when viewed face-on ($\mu^{\rm fo}$) is often parametrized as:
\begin{equation}
\mu(B)  \equiv \mu\Bigl({\rm cos}(i)\Bigr)= \mu^{\rm fo} + C\times2.5\,{\rm log}\Bigl({\rm cos}(i)\Bigr)~.
\label{eq:fitopac}
\end{equation}
Here $i$ is the inclination angle ($i$\,=\,0 for face-on and $\nicefrac{\pi}{2}$ for edge-on systems). The amount of attenuation is parametrized by the opacity constant $C$ which ranges from unity for transparent disks to zero for opaque disks. For a non-zero $C$, the apparent surface brightness of an inclined disk is thus always more intense than in the face-on object.

\begin{figure}[ht]
\centering
\includegraphics[scale=0.65, angle=-90]{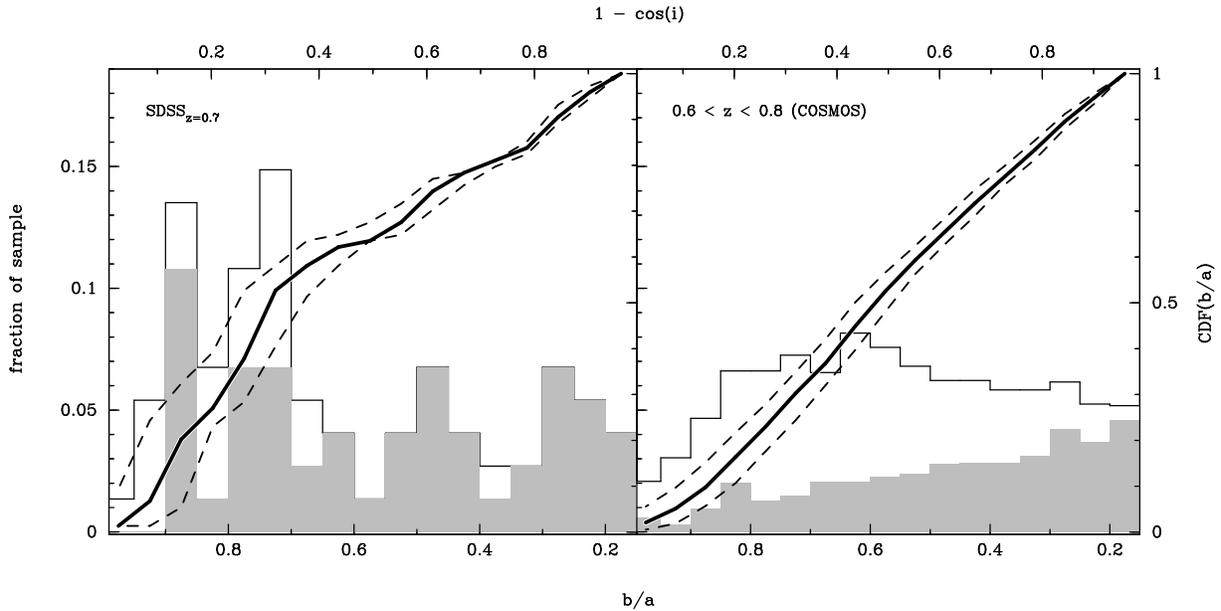}
\caption{Distribution of axis ratios $b/a$ for pure disk galaxies at $0.6$\,$\leq$\,$z$\,$<$\,0.8 in the COSMOS field ({\it right}\,) and SDSS$_{z=0.7}$ disks ({\it left}\,). The histograms are normalized to the total number of galaxies in the two samples. The distributions of axis ratios of disks with $r_{1/2}$\,$\geq$\,5\,kpc are indicated in grey. The upper axis shows the correspondence between inclination $i$ and axis ratio $b/a$.\newline
The cumulative distribution function (CDF) of axis ratios in the total disk population is plotted with a solid black line (see scale at right). Dashed lines delimit the associated 99\% confidence region. \label{fig:axratincls}}
\end{figure}

\noindent The distribution of axis ratios $b/a$ can be an indication of the opacity of disks \citep[e.g.][]{jones96}. In Fig. \ref{fig:axratincls} we plot the axis ratios for our COSMOS disks ({\it right}\,) and the SDSS$_{z=0.7}$ sample ({\it left}\,). Instances of $b/a$\,$\sim$\,1 and 0 are rare due to intrinsic disk ellipticity \citep{ryden04} and the finite thickness of edge-on systems. A one-sided Kolmogorov-Smirnov (K-S) test indicates that -- within the errors shown -- there is a non-negligible probability that the measured axis ratios in the entire samples are drawn from a flat distribution (SDSS$_{z=0.7}$: $p$\,=\,0.19; COSMOS: $p$\,=\,0.16), as illustrated by the cumulative distribution function crossing both panels on an approximate diagonal. Fig. \ref{fig:axratincls} also shows the histograms (in grey) of $b/a$ for the subset of disks with physical half-light radii $r_{1/2}$\,$\geq$\,5\,kpc. We introduce this restriction in order to obtain a {\it complete}\footnote{~The threshold of 5\,kpc corresponds to the minimal size at $z$\,$\sim$\,0.7 which ensures that the selection limit in magnitude ($I$\,=\,22.5) can be reached for the entire range of `normal' surface brightness values (i.e. neglecting likely rare very low surface brightness disks), cf. \citet{lilly98} and \citet{sargent07}.} sample of disks for the final analysis. While the axis ratio distribution of large SDSS$_{z=0.7}$ disks remains flat (K-S probability $p$\,=\,0.99), that of the corresponding COSMOS population is skewed toward elongated objects ($p$\,$<$\,$10^{-5}$ for the distribution being flat). The different behaviour hints at opacity variations which we quantify in the next paragraph.

\begin{figure}[ht]
\centering
\includegraphics[scale=0.65, angle=-90]{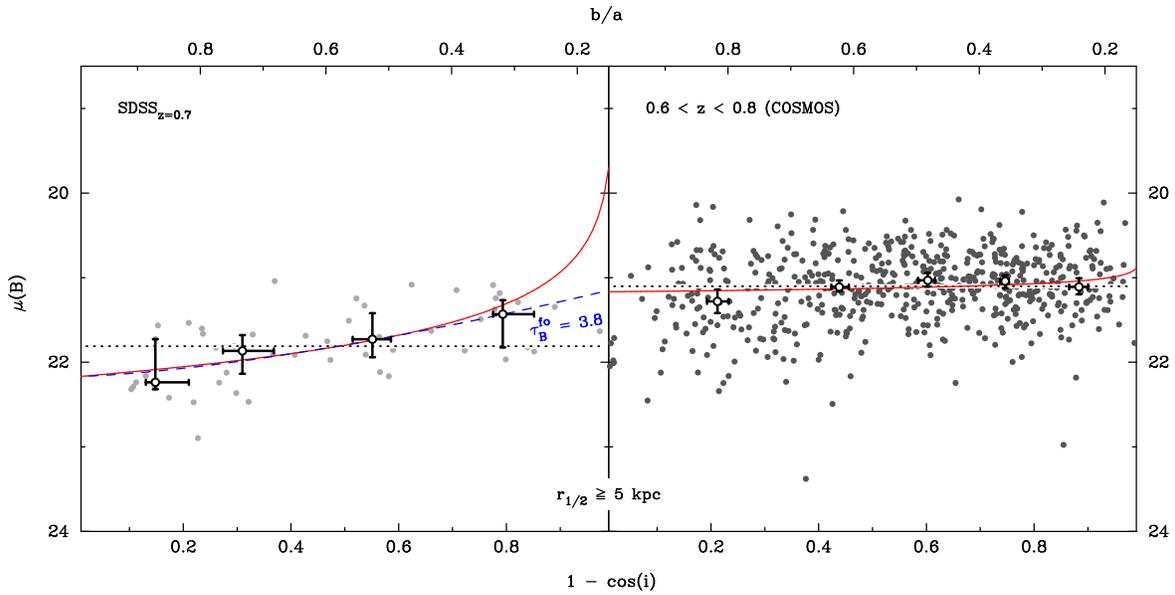}
\caption{Rest frame $B$-band surface brightness-inclination relation for large ($r_{1/2}\geq5$\,kpc) artificially redshifted local disk galaxies ({\it left}\,) and observed large disks at $z$\,$\sim$\,0.7 ({\it right}\,). Dotted lines: median surface brightness averaged over all inclinations. Black circles: median surface brightness and inclination in discrete bins of inclination (error bars span the 95\% confidence interval). Red lines represent the best-fitting surface brightness-inclination relation in eq. (\ref{eq:fitopac}) for local and $z$\,$\sim$\,0.7 disks. On the left, the blue line shows the prediction by \citet{moellenhoff06} based on the mean $B$-band opacity of local spiral disks in the Millenium Galaxy Catalogue \citep[$\tau_B^{\rm fo}$\,=\,3.8$\pm$0.7;][]{driver07}. \label{fig:muincl}} 
\end{figure}

\noindent Axis ratios are related to inclination (cf. also scale along upper edge of Fig. \ref{fig:axratincls}) by the \citet{hubble26} formula:
\begin{equation}
{\rm cos}^2(i) = \frac{(\nicefrac{b}{a})^2-(\nicefrac{b}{a})^2_{\rm min}}{1-(\nicefrac{b}{a})^2_{\rm min}}~,
\label{eq:cosiq}
\end{equation}
with $(b/a)_{\rm min}$\,=\,0.15, in keeping with measurements of the minimal flattening in late-type spirals \citep[e.g.][]{guthrie92, yuan04}.\\
We present the surface brightness-inclination relation for large SDSS$_{z=0.7}$ and COSMOS disks (51 and 611 objects, respectively) in Fig. \ref{fig:muincl}. The error bars on the median surface brightness values in bins of increasing inclination (cf. also Table \ref{tab:meds}) span the 95\% confidence region, estimated with $N_{{\rm cos}(i)}\times$\,100 bootstrap realizations ($N_{{\rm cos}(i)}$ is the number of disks in a given bin of inclination). The uncertainties on the COSMOS median are significantly smaller, in agreement with the expectation that they should scale roughly as $\nicefrac{1}{\sqrt{N_{{\rm cos}(i)}}}$.\\
The dashed blue line in Fig. \ref{fig:muincl} ({\it left}\,) marks the surface brightness-inclination relation for a $B$-band central face-on optical depth of $\tau_B^{\rm fo}$\,=\,3.8, the average opacity of nearby disks recently determined by \citet{driver07}. The plotted line is taken from the predictions\footnote{~The fact that our pure disks are fit by a single S\'ersic component implies that average and central surface brightness $\mu_0(B)$ are related by a constant offset \citep[see, e.g.,][]{grahamdriver05}. The predictions of \citet{moellenhoff06} for the inclination-dependence of $\overline{\mu_0}(B)$ can thus be meaningfully compared to our findings.}  in \citet{moellenhoff06}, based on radiative transfer models of stable local disks \citep{popescu00, tuffs04}. The good agreement with our measurements demonstrates that we correctly recover the attenuation properties of local disks even when they are redshifted to $z$\,$\sim$\,0.7.

\begin{figure}[ht]
\centering
\includegraphics[scale=0.75, angle=-90]{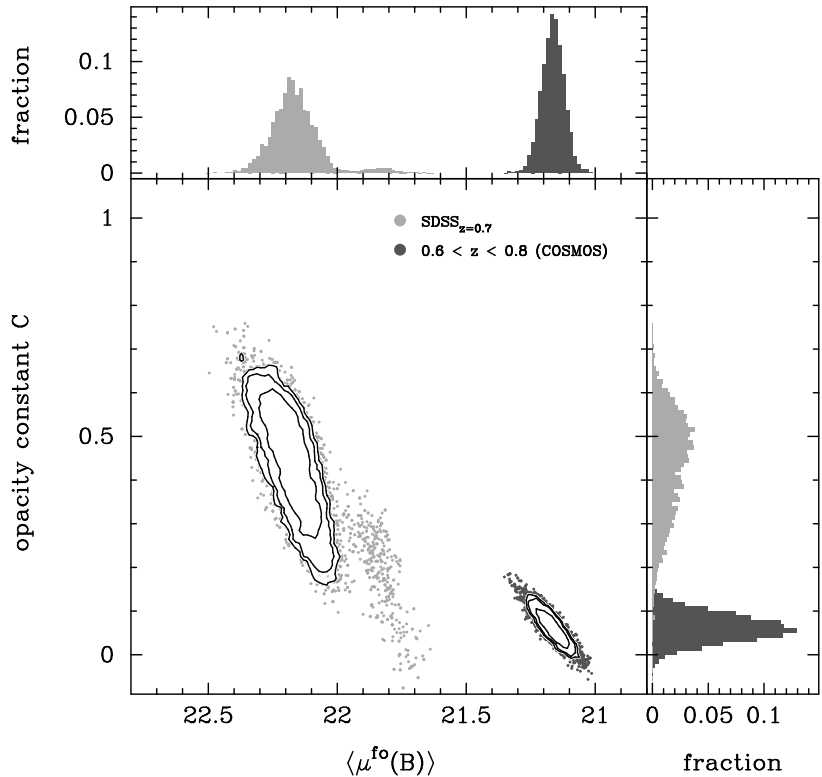}
\caption{Distribution of the best-fit parameter pairs ($\langle\mu^{\rm fo}(B)\rangle$,\,$C$) in eq. (\ref{eq:fitopac}), obtained by repeated fits to the resampled data in Table \ref{tab:meds} (see text for details). Light and dark grey is used for the low and high redshift sample, respectively, as in Fig. \ref{fig:muincl}. The contours in the main window are isopycnics enclosing 95, 90 and 68\% of the points. The panels along the edges of the figure show the projected distributions of average face-on surface brightness in the $B$-band, $\langle\mu^{\rm fo}(B)\rangle$ ({\it top}\,), and of the opacity constant $C$ ({\it right}). \label{fig:finpar}} 
\end{figure}

\noindent We now determine the average correction between face-on and observed surface brightness. At first sight, Fig. \ref{fig:muincl} might suggest that -- within uncertainties -- the surface brightness-inclination relation is flat both for SDSS$_{z=0.7}$ and COSMOS disks (see dotted horizontal line marking the global surface brightness average). However, as we will now show, the opacity constants describing local and $z$\,$\sim$\,0.7 disks turn out to be significantly different.\\
To properly account for strongly asymmetric error bars (see Fig. \ref{fig:muincl}) we interpret the distribution of bootstrapped medians as a probability distribution and accordingly draw pairs of medians ($\overline{{\rm cos}(i)}$, $\overline{\mu}(B)$) in all bins of inclination. We then fit eq. (\ref{eq:fitopac}) to each of the sets of resampled medians and subsequently locate the peak of the resulting distributions of the free parameters $C$ and $\mu^{\rm fo}$.\\
Fig. \ref{fig:finpar} (main window) shows the best-fitting parameter pairs ($\langle\mu^{\rm fo}(B)\rangle$,\,$C$) from 10,000 realizations. The cores of the point clouds are highlighted with (smoothed) equal density contours that enclose 68, 90 and 95\% of the points. By projecting the scatter plots along the vertical and horizontal axes we obtain 95\% confidence intervals\footnote{~Note that the 95\% confidence limits include the diffuse cloud of outliers at $\langle\mu^{\rm fo}(B)\rangle$\,$\approx$\,21.8 (Fig. \ref{fig:finpar}), occurring as a result of small number statistics in the first and last SDSS$_{z=0.7}$ inclination bin.} for the free parameters $\langle\mu^{\rm fo}(B)\rangle$ and $C$ for the local and high-$z$ disk samples (Table \ref{tab:finpar}). The average surface brightness of face-on disks, $\langle\mu^{\rm fo}(B)\rangle$, increases by $\sim$\,1\,mag between $z$\,$\sim$\,0 and 0.7 \citep[cf. also][]{lilly98, barden05}. For the opacity constant we find $C$($z$\,$\sim$\,0.7)\,$\in$\,[0.18,\,0.62] and $C$($z$\,$\sim$\,0)\,$\in$\,[0.01,\,0.13] and for the most probable values\footnote{~The evolution of the opacity constant $C$ is not brought about by differing distributions of axis ratios in the two samples. Resampling the surface brightness values measured for large COSMOS disks according to the distribution of $b/a$ in the SDSS$_{z=0.7}$ sample also results in a flat surface brightness-inclination relation at $z$\,$\sim$\,0.7.} $C$($z$\,$\sim$\,0)\,=\,0.47 and $C$($z$\,$\sim$\,0.7)\,=\,0.07. The corresponding surface brightness-inclination dependence (eq. \ref{eq:fitopac}) is indicated in red in Fig. \ref{fig:muincl}.\\
We note that by using the `unbrightened' sample of \citet{kampczyk07} (see \S\,\ref{sec:SDSS}) only the most luminous local disks satisfy the selection criterion $I$\,$\leq$\,22.5 when redshifted to {\it z}\,$\sim$\,0.7. The general luminosity evolution of the disk galaxy population \citep[e.g.][]{scarlata07} implies that only the brightest COSMOS disks at 0.6\,$<$\,{\it z}\,$<$\,0.8 have masses similar to those of the SDSS$_{z=0.7}$ sample. As the degree of extinction depends on the amount of attenuating material \citep[e.g.][]{masters10}, this could be partly responsible for the derived difference in the opacity constant $C$. We can generate a sample of COSMOS disks that, to first order, have masses similar to the SDSS$_{z=0.7}$ galaxies by brightening the selection limit for COSMOS galaxies by $\sim$${\Delta}z$ magnitudes to $I$\,$\approx$\,21.8. This restriction does not change our finding of a nearly flat ($C$($z$\,$\sim$\,0.7)\,=\,0.05$_{-0.14}^{+0.08}$, in this case) surface brightness-inclination relation at {\it z}\,$\sim$\,0.7.

\section{Conclusions}
\label{sec:conclu}
We have investigated the average opacity of distant (0.6\,$<$\,$z$\,$<$\,0.8) COSMOS spiral galaxies by direct comparison with local disks artificially redshifted to $z$\,$\sim$\,0.7. The outcome of inclination-based attenuation measurements is susceptible to selection effects \citep[e.g.][]{davies93, jones96}. By processing our two samples identically and by applying a morphologically and photometrically consistent sample selection, we have taken the necessary precautions to reduce such systematic biases. We recover the expected surface brightness-inclination relation for nearby disks even after the simulated shift to {\it z}\,$\sim$\,0.7 (in the parameterization of eq. (\ref{eq:fitopac}) this implies an opacity constant $C$($z$\,$\sim$\,0)\,=\,0.47$_{-0.28}^{+0.16}$). For high-$z$ COSMOS disks a significantly lower value of $C$($z$\,$\sim$\,0.7)\,=\,0.07$\pm$0.06 is found, implying, on average, a nearly constant relation between surface brightness and inclination as is expected for optically thick spiral galaxies.

\noindent Previous studies suggest that the extinction laws in star-forming galaxies at similar redshifts as our COSMOS sample do not differ strongly from those in local systems (e.g., \citealp{conroy09} or \citealp{calzetti01} and references therein). It thus seems unlikely that the increased opacity of our {\it z}\,$\sim$\,0.7 COSMOS disks is due to a different chemical composition of the dust. Given that our low- and high-{\it z} disk samples have similar luminosities we can also rule out that the evolution is due to the locally observed scaling of dust opacity with the blue luminosity of galaxies \citep{wangheckman96}.\\
Other possible explanations for the flat surface brightness-inclination relation at $z$\,$\sim$\,0.7 are the presence of more attenuating material, or a different spatial arrangement thereof. Evidence that both factors might contribute exists: \citet{genzel08} report stronger turbulent motion in disk-like systems at high redshift that could increase the scale height of the dust, and recent measurements of molecular line emission in typical late type galaxies at $z$\,$\sim$\,1.5 have revealed large gas fractions in excess of 50\% of the baryonic mass \citep{daddi09}. The measurements presented here are not sufficient to infer the relative importance of such potential contributions. Additional constraints from complementary measurements or different wavelength regions are thus indispensable to determine the causes for the opacity evolution we observe between {\it z}\,$\sim$\,0 and 0.7.

\acknowledgments
We gratefully acknowledge the anonymous referee's helpful suggestions and the contribution of the COSMOS collaboration and its more than 100 scientists worldwide. PK, PAO, CS and MTS were supported by the Swiss National Science Foundation. This research was also financed by DFG grant SCHI 536/3-2. The HST COSMOS Treasury program was supported through NASA grant HST-GO-09822.

\noindent {\it Facilities:} \facility{HST (ACS)}, \facility{Subaru (SuprimeCam)}

\clearpage

\begin{deluxetable}{lcc}
\tabletypesize{\scriptsize}
\tablecaption{Median values of inclination (col. 2) and of the associated average rest-frame $B$-band surface brightness (col. 3; cf. also Fig.\,\ref{fig:muincl}). Errors span the 95\% confidence interval. \label{tab:meds}}
\tablewidth{0pt}
\setlength{\tabcolsep}{.15in}
\tablehead{
\colhead{sample/redshift} &
\colhead{$1-\overline{{\rm cos}(i)}$} &
\colhead{$\overline{\mu}(B)$} }
\startdata
SDSS$_{z=0.7}$ & 0.148$_{-0.018}^{+0.062}$ & 22.237$_{-0.511}^{+0.082}$\\[1ex]
& 0.310$_{-0.036}^{+0.058}$ & 21.866$_{-0.187}^{+0.269}$\\[1ex]
& 0.551$_{-0.037}^{+0.034}$ & 21.728$_{-0.307}^{+0.213}$\\[1ex]
& 0.793$_{-0.018}^{+0.058}$ & 21.432$_{-0.165}^{+0.393}$\\[1ex]
\hline\\[-2ex]
COSMOS & 0.212$_{-0.019}^{+0.021}$ & 21.281$_{-0.142}^{+0.138}$\\[1ex]
\raisebox{1ex}[1ex]{($z$\,$\sim$\,0.7)} & 0.438$_{-0.015}^{+0.018}$ & 21.110$_{-0.078}^{+0.054}$\\[1ex]
& 0.603$_{-0.018}^{+0.013}$ & 21.031$_{-0.087}^{+0.073}$\\[1ex]
& 0.746$_{-0.009}^{+0.007}$ & 21.042$_{-0.070}^{+0.082}$\\[1ex]
& 0.883$_{-0.018}^{+0.017}$ & 21.109$_{-0.102}^{+0.089}$
\enddata
\end{deluxetable}

\begin{deluxetable}{lcc}
\tabletypesize{\scriptsize}
\tablecaption{Favoured values (cf. Fig. \ref{fig:finpar}) of the free parameters $\langle\mu^{\rm fo}(B)\rangle$ and $C$ in the surface brightness-inclination relation of eq. (\ref{eq:fitopac}). Errors span the 95\% confidence interval. \label{tab:finpar}}
\tablewidth{0pt}
\setlength{\tabcolsep}{.15in}
\tablehead{
\colhead{sample/redshift} &
\colhead{$\langle\mu^{\rm fo}(B)\rangle$} &
\colhead{$C$} }
\startdata
SDSS$_{z=0.7}$ & 22.172$_{-0.339}^{+0.158}$ & 0.465$_{-0.281}^{+0.158}$\\[2ex]
COSMOS ($z$\,$\sim$\,0.7) & 21.168$_{-0.086}^{+0.091}$ & 0.065$_{-0.061}^{+0.060}$
\enddata
\end{deluxetable}

\quad\\

 \end{document}